\documentstyle[12pt]{article}
\begin{document}

%\SetTitle{A simple article}
%\SetAuthor{A. U. Thor}
%\Setdate{}
%\TitlePage{}

\LaTeX{}\bigskip

\begin{center}
The plasma-solid transition:two simple examples\smallskip \medskip

V.Celebonovic$^{1}$ and W.Dappen$^{2}$

$^{1}$Institute of Physics,Pregrevica 118,11080 Zemun-Beograd,Yugoslavia

$^{2}$Dept.of Physics and Astronomy,USC,Los Angeles CA 90089-1342,USA\bigskip

1. Introduction\medskip
\end{center}

The notion of ''astrophysical matter'' is meant by most people to refer to
one of the various forms of plasmas.However,this is not entirely
true,because contemporary astrophysics is full of cases of objects built
with dominantly high-pressure,solid-state matter.Obvious examples of such
dense objects are planetary and stellar interiors $\left[ 1\right] $.
Logically then,between these two regimes,
there must exist a region in which a plasma-solid (PS) transition occurs.

The aim of this contribution is to explore in some detail the conditions for
the PS transition,using a simple criterion as proposed in $\left[ 2\right] $
.We apply it to the two extreme cases: Fermi-Dirac (FD) and Bose-Einstein
(BE) gas.One might object here that the applicability of any of these model
systems to real astrophysical objects is not very probable.This is true,but
the idea behind the present calculation is to obtain an order-of-magnitude
estimate,first in the two pure limits (FD and BE),and later in a mixed FD
and BE case.\medskip

\begin{center}
2. Calculations
\end{center}

The criterion for the existence of the solid phase, proposed in $\left[
2\right] $, has the form:

\begin{equation}
T\cdot 10^{4}T_{6}^{2}\leq \rho \leq 3\cdot 10^{11}Z^{10}  \label{(1)}
\end{equation}

and 
\begin{equation}
50\cdot Z^{-5}T_{6}^{3}\leq \rho \leq 7\cdot 10^{4}T_{6}^{2}  \label{(2)}
\end{equation}

where $T_{6}=T/10^{6}$ and $\rho $ is the mass density.Eq.(1) is applicable
to ''low temperatures''.\newpage

\begin{center}
2.1 The FD case
\end{center}

The equation of state (EOS) of a FD gas has the form:

\begin{equation}
n=\frac{4\pi }{h^{3}}(2mT)^{3/2}F_{1/2}\left( \frac{\mu }{T}\right)
\label{(3)}
\end{equation}

where $k_{B}=1,\mu $ is the chemical potential,and $F_{1/2}$ is the FD
integral of the order $1/2$ $\left[ 3\right] $ .

According to $\left[ 3\right] $

\begin{equation}
F_{1/2}\left( \frac{\mu }{T}\right) \simeq \frac{2}{3}\mu _{0}^{3}\left[ 1+%
\frac{\pi ^{4}}{48}\left( \frac{T}{\mu _{0}}\right) ^{4}\right]  \label{(4)}
\end{equation}

where $\mu _{0}=Cn^{2/3}$ and $C=\left( 3\pi ^{2}\right) ^{2/3}\frac{\hbar
^{2}}{2m}$ . Inserting eq.(4) into eq.(3),it follows that

\begin{equation}
n\simeq \left( \frac{\pi ^{4}}{48}\right) ^{3/8}\left( \frac{T}{C}\right)
^{3/2}\left( \frac{T^{3/2}}{8}\right) ^{3/8}\left( \frac{T^{3/2}}{8\left( 1-%
\frac{T^{3/2}}{8}\right) }\right) ^{3/8}  \label{(5)}
\end{equation}

The mass density $\rho $ and the electron number density $n$ are related by $%
n=N_{A}\frac{Z\rho }{A}$.Applying this relation and eq.(5) to eq.(1),one
finally gets 
\begin{equation}
10^{-8}T^{3}\leq \frac{A}{N_{A}Z}\frac{1}{864^{1/8}4}\left( \frac{\pi }{C}%
\right) ^{3/2}\left( T^{33/16}+\frac{3}{64}T^{57/16}\right) \leq 3\cdot
10^{11}Z^{10}  \label{(6)}
\end{equation}

which is the allowed interval for the existence of the solid phase of a FD
gas.

\begin{center}
2.2 The BE case
\end{center}

The EOS of an ideal BE gas has the form $\left[ 4\right] $

\begin{equation}
n=\frac{g}{\hbar ^{3}}\left( \frac{mT}{2\pi }\right) ^{3/2}Li_{3/2}\left[
\exp \left( \frac{\mu }{T}\right) \right]  \label{(7)}
\end{equation}

where $g=2s+1$ is the spin degeneracy factor and $Li_{n}\left[ z\right]
=\sum_{k=1}^{\infty }\frac{z^{k}}{k^{n}}$ is the definition of the
polylogarithm.One of the important points of the phase diagram of a BE gas
is the Bose-Einstein condensation (BEC) point,defined as the zero point of
the chemical potential.
\newpage
Inserting this condition into eq.(7),and performing the necessary algebra,one 
gets the following result for the BEC temperature

\begin{equation}
T_{BEC}=\frac{2\pi \hbar ^{2}}{\left[ g\zeta \left( 3/2\right) \right] ^{2/3}%
}\frac{n^{2/3}}{m}  \label{(8)}
\end{equation}

($n$ - the number density of a BE gas ; $\zeta $ - the Riemann zeta
function).Solving eq.(8) for the number density and inserting the result
into eq.(1), it follows that

\begin{equation}
\frac{2\pi \hbar ^{2}}{(10^{16}m^{5})^{1/3}}\frac{T^{2}}{\left[ g\zeta
\left( 3/2\right) \right] ^{2/3}}\leq T_{BEC}\leq 2\pi \hbar ^{2}\left( 
\frac{3\cdot 10^{11}}{g\zeta \left( 3/2\right) }\right) ^{2/3}\frac{1}{%
m^{5/3}}  \label{(9)}
\end{equation}

This inequality gives the interval of temperatures in which the BEC
temperature has to be in order for the BEC to occur in a solid phase.In
deriving it,we have used $\rho =mn$.

\begin{center}
3. Discussion and conclusions
\end{center}

In this contribution we have discussed the plasma-solid traansition for two
model systems: the FD and the BE gas.The calculations were performed
according to a criterion proposed by Kirzhnitz,and the results are given in
eqs.(6) and (9).These two expressions are to be considered as limiting cases
of astrophysical reality.Of cours,in the most common astrophysical
systems,such as stellar interiors and atmospheres,one encounters mixture of
the general form BE and FD.The problem of a PS transition in such systems
will be treated elsewhere.

However,our eq.(9) is applicable in the exotic albeit real astrophysical
setting of white dwarfs and neutron stars (and similar other dense
objects).There are indications that a PS transition might occur in them $%
\left[ 5\right] .$Comparing the results of eq.(9) with the temperature
profiles of neutron stars,in the domain after the PS transition,one could
determine the depth at which BEC could possibly occur in these objects.This
is just a qualitative indication of the possible applicability of eq.(9),and
the details will be considered at length elsewhere.
\newpage
\begin{center}
References
\end{center}

$\left[ 1\right] $ E.Schatzman and F.Praderie: Astrophysique:Les
Etoiles,
InterEditions/Editions du CNRS,Paris (1990).

$\left[ 2\right] $ D.A.Kirzhnitz: Zh.Exp.Theor.Fiz.,{\bf 38},503 (1960).

$\left[ 3\right] $ V.Celebonovic: Publ.Astron.Obs.Belgrade,{\bf 60},16 (1998).

$\left[ 4\right] $ L.E.Reichl: A Modern Course in Statistical Physics,Edward
Arnold (Publishers) Ltd.,London (1988).

$\left[ 5\right] $ D.G.Yakovlev,K.P.Levenfish and Yu.A.Shibanov: Cooling
Neutron stars and Superfluidity in Their Interiors,preprint astro-ph/9906456
(1999).

\end{document}